\documentclass[aps,pra,reprint,superscriptaddress]{revtex4-1} %aps,prl,reprint
\usepackage{graphicx}
\usepackage[colorlinks,citecolor=blue,urlcolor=blue,linkcolor=blue]{hyperref}
\usepackage{amsmath}
\usepackage{braket}
\usepackage{booktabs}
\usepackage[squaren]{SIunits}
\usepackage{booktabs}
\usepackage{supertabular}
\usepackage{multirow}
\usepackage{rotating}
\usepackage{widetable}
\usepackage{bm}

\begin{document}

\title{Temporally versatile polarization entanglement from Bragg-reflection waveguides}

\author{A. Schlager}
\affiliation{Institut f\"ur Experimentalphysik, Universit\"at Innsbruck, Technikerstra\ss e 25, 6020 Innsbruck, Austria}

\author{B. Pressl}
\affiliation{Institut f\"ur Experimentalphysik, Universit\"at Innsbruck, Technikerstra\ss e 25, 6020 Innsbruck, Austria}

\author{K. Laiho}
\email{kaisa.laiho@uibk.ac.at}
\affiliation{Institut f\"ur Experimentalphysik, Universit\"at Innsbruck, Technikerstra\ss e 25, 6020 Innsbruck, Austria}

\author{H. Suchomel }
\affiliation{Technische Physik, Universit\"at W\"urzburg, Am Hubland,  97074 W\"urzburg, Germany}

\author{M. Kamp}
\affiliation{Technische Physik, Universit\"at W\"urzburg, Am Hubland,  97074 W\"urzburg, Germany}

\author{S. H\"ofling}
\affiliation{Technische Physik, Universit\"at W\"urzburg, Am Hubland,  97074 W\"urzburg, Germany}
\affiliation{School of Physics $\&$ Astronomy, University of St Andrews, St Andrews, KY16 9SS, United~Kingdom}

\author{C. Schneider}
\affiliation{Technische Physik, Universit\"at W\"urzburg, Am Hubland,  97074 W\"urzburg, Germany}

\author{G. Weihs}
\affiliation{Institut f\"ur Experimentalphysik, Universit\"at Innsbruck, Technikerstra\ss e 25, 6020 Innsbruck, Austria}

\begin{abstract}
Bragg-reflection waveguides emitting broadband parametric down-conversion (PDC) have been proven to be well suited for the on-chip generation of polarization entanglement in a straightforward fashion [R. T. Horn et al., Sci. Rep. 3, 2314 (2013)]. Here, we investigate how the properties of the created states can be modified by controlling the relative temporal delay between the pair of photons created via PDC. Our results  offer an easily accessible approach for changing the coherence of the polarization entanglement, in other words, to tune the phase of the off-diagonal elements of the density matrix. Furthermore, we provide valuable insight in the engineering of these states directly at the source. 
\end{abstract}

\maketitle

%----------------------------------------------------------------------------------
%                            Introduction
%----------------------------------------------------------------------------------

The verification of the non-classical nature of quantum objects, such as photons, plays an important role to ensure a successful implementation of forthcoming quantum technologies. Among theses tasks is the preparation and characterization of polarization entanglement that provides some of the strongest evidence of the quantum features of light \cite{Edamatsu2007, Pan2012, Alibart2016}. Since the first realizations with parametric down-conversion (PDC) in bulk crystals more than two decades ago  \cite{Ou1988, Shih1988, Kwiat1995}, a lot of effort has been put in replacing them both with ferroelectric \cite{Jiang2006, Lim2008, Zhong2010, Martin2010,  Herrmann2013} and semiconductor waveguides \cite{Valles2013, Orieux2013,Horn2013}, which provide higher brightnesses and easier alignment in collinear configurations.

In one of such configurations, the cross-polarized PDC photon pairs--called signal and idler--that have at least some nanometers of spectral extent, are divided on a dichroic beam splitter into two paths \cite{Levine2011, Kaiser2012, Bruno2014, Kuo2016,Laudenbach2016}. Due to a strict frequency or energy correlation between signal and idler, a two-partite polarization superposition is created between the output paths of the dichroic mirror. Often, a continuous-wave pump laser is utilized to guarantee tight correlations in the spectral domain \cite{Horn2013, Kang2015}. 

Bragg-reflection waveguides (BRWs) made of AlGaAs generate indistinguishable photon pairs over several tens of nanometers \cite{Horn2013, Guenthner2015} and they have been utilized for narrowband multiplexing of polarization entanglement \cite{Kang2016, Autebert2016a}. Due to the very small birefringence between the orthogonally-polarized signal and idler \cite{Laiho2016}, BRWs may work reasonably well in this configuration even without compensating their temporal walk-off. However, their group index difference is large enough to result into an uncompensated phase. Therefore, the engineering of the spectro-temporal degree of freedom of PDC photon pairs from BRWs is essential for the controlled creation and manipulation of the polarization entanglement.
 
Changes in the coherence of the polarization entangled states,  that is their phase,  provide an interesting degree of freedom for the state manipulation \cite{Pan2012}. When regarding PDC sources, this is usually accomplished by modifying the characteristics of one of the  entangled parties with birefringent retarders \cite{Mattle1996, Sansoni2010, Horn2013}. However, the phase control over the joint state can possibly be realized  more easily \cite{Kwiat1999, Steinlechner2012, Chen2015}.
Here, we generate broadband, pulsed polarization entanglement from a collinear BRW source emitting PDC.  As the intrinsic group delay of signal and idler causes an uncompensated phase, we  tailor the coherence of the created state by introducing an optical path length difference for them.  Our results show that the phase of the created state can be tuned flexibly in a very simple manner. Further, we provide an important insight into how the spectro-temporal properties of the joint PDC state can be taken advantage of in the creation and manipulation of polarization entanglement.

%----------------------------------------------------------------------------------
%                        Theory
%----------------------------------------------------------------------------------

Our BRW source generates cross-polarized signal and idler photons via a type-II PDC process. In the low gain regime, the photon-pair state can be approximated as \cite{W.P.Grice2001, Y.-H.Kim2004}
\begin{equation}
\label{eq_psi}
\ket{\psi} = \int \hspace{-1.5ex} \int d\omega_{s} d \omega_{i} f(\omega_{s}, \omega_{i})\hat{a}_{H}^{\dagger}(\omega_s)\hat{a}^{\dagger}_{V}(\omega_{i})\ket{0},
\end{equation}
in which $f(\omega_{s}, \omega_{i})$ describes the joint spectral amplitude (JSA) of signal ($s$) and idler ($i$) in terms of their angular frequencies $\omega_{\mu}$ ($\mu = s,i$). The operator $\hat{a}_{\sigma}^{\dagger}(\omega)$ accounts for the  creation of a photon with a specific angular frequency at either horizontal ($H$) or vertical ($V$) polarization ($\sigma = H,V$). The normalization condition ensures that $\int \hspace{-1ex} \int d\omega_{s} d \omega_{i} |f(\omega_{s}, \omega_{i})|^{2} = 1$.

\begin{figure}
\centering
\includegraphics[width = 0.44\textwidth]{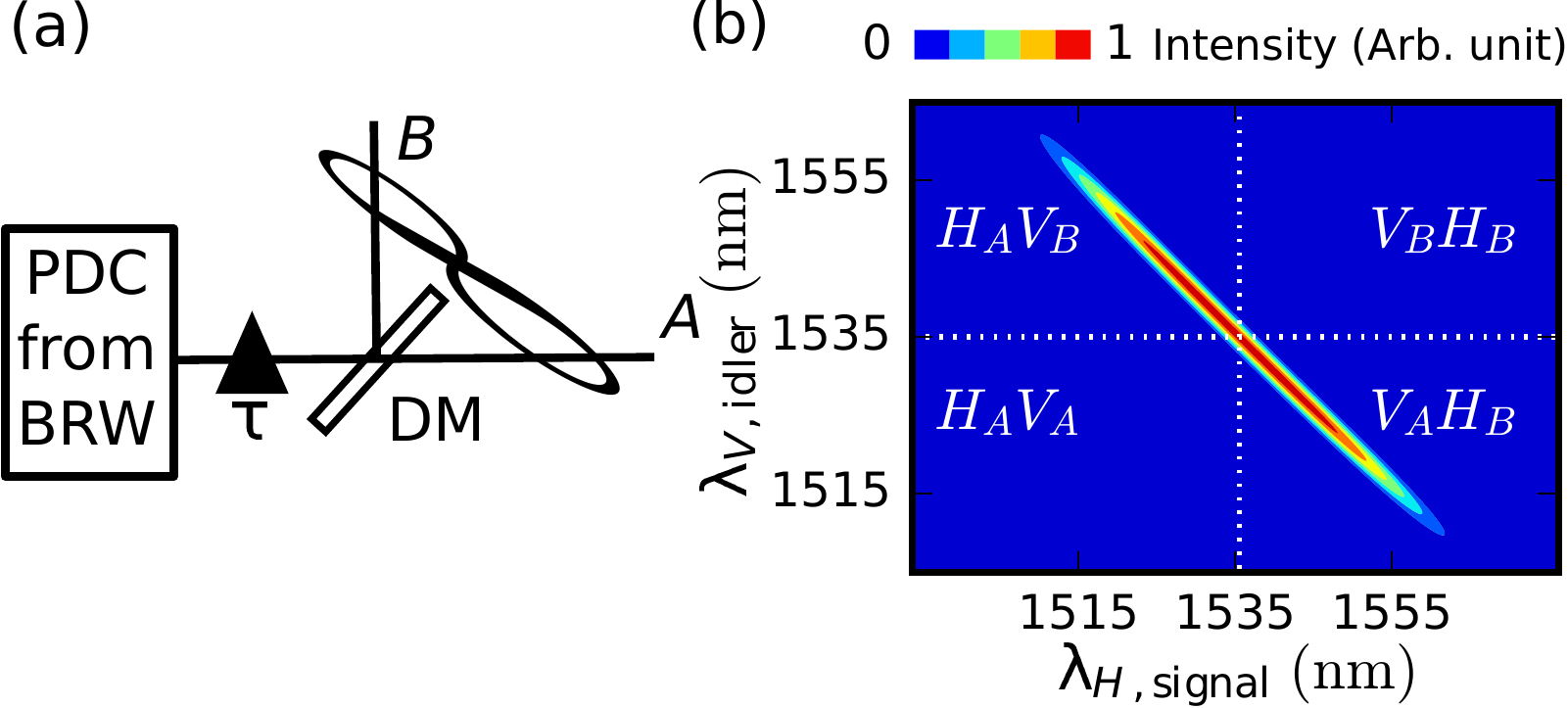}
\caption{\label{fig:polent} (a) Scheme for creating polarization entangled states via the type-II PDC emission from a BRW, in which signal and idler pass a polarization selective delay of $\tau$ before being split at a dichroic mirror (DM) into two paths $A$ and $B$. (b) The absolute value of the JSA for our BRW in terms of the wavelength $\lambda$ subjected to the spectral filtering of our experiment and  calculated using the PDC process parameters measured in Refs~\cite{B.Pressl2015,Laiho2016}. The output paths taken by the $H$-polarized signal and $V$-polarized idler are shown in the different spectral regions. The dotted lines mark the cut-off edge of the DM.}
\end{figure}

For the creation of polarization entanglement we employ the  scheme illustrated in Fig.~\ref{fig:polent}(a), also used for BRWs in Refs~\cite{Horn2013, Kang2015} without delay compensation. The state from \eqref{eq_psi} is split at a dichroic mirror, which we model as a frequency dependent beam splitter. We take into account the finite steepness of the cut-off edge of the dichroic beam splitter and its polarization dependent behavior. Prior to that, the signal beam is delayed by a time $\tau$ with respect to the idler. Thus, we apply the beam splitter transformations
\begin{align}
\label{eq_BS}
&\hat{a}_{H}^{\dagger}(\omega_s)=  \left [ \sqrt{T_{H}(\omega_s)}\hat{A}_{H}^{\dagger}(\omega_s) -\sqrt{R_{H}(\omega_s) }
\hat{B}_{H}^{\dagger}(\omega_s) \right ] \textrm{e}^{-i\omega_s \tau} \quad{\textrm{and}} \nonumber\\
&\hat{a}_{V}^{\dagger}(\omega_i) = \sqrt{ T_{V}(\omega_i)}\hat{A}_{V}^{\dagger}(\omega_i) -\sqrt{R_{V}(\omega_i) }\hat{B}_{V}^{\dagger}(\omega_i),
\end{align}
in which  $T_{\sigma}(\omega_{\mu})$ and $R_{\sigma}(\omega_{\mu})$ are the  polarization dependent, real-valued transmission and  reflection coefficients of the dichroic mirror, for which the relation $T_{\sigma}(\omega_{\mu})+R_{\sigma}(\omega_{\mu})= 1$ holds. Further, in \eqref{eq_BS}~the properties of signal and idler are expressed in terms of the transmitted and reflected output arms of the dichroic mirror labeled with $A$ and $B$, respectively. 

We plug the transformations in \eqref{eq_BS} into \eqref{eq_psi} in order to derive the polarization entangled state. We keep only the terms that can cause coincidences between $A$ and $B$. This approximation is justified by the shape of the JSA in our BRW and only a very small part of the JSA is  neglected as illustrated in Fig.~\ref{fig:polent}(b). We express the post-selected state in the form
\begin{align}
\label{eq_PP}
\ket{\phi} \approx \int \hspace{-1ex} \int d\omega_{s} d \omega_{i}
\textrm{e}^{-i\omega_s \tau} \Big [ &g(\omega_{s}, \omega_{i})\ket{\omega_{s}, H}_{A}\ket{\omega_{i}, V}_{B} \nonumber \\
 + &h(\omega_{s}, \omega_{i})\ket{\omega_{i}, V}_{A}\ket{\omega_{s}, H}_{B} \Big],
\end{align}
in which 
\begin{equation}
g(\omega_s,\omega_i) = f(\omega_s,\omega_i)\sqrt{T_H(\omega_s)R_V(\omega_i)}  \quad{\textrm{and}}
\end{equation}
\begin{equation}
h(\omega_s,\omega_i) = f(\omega_s,\omega_i)\sqrt{R_H(\omega_s)T_V(\omega_i)}.
\end{equation}
Further, we use in \eqref{eq_PP} the short notation $\hat{A}_{\sigma}^{\dagger}(\omega_{\mu})\ket{0} = \ket{\omega_{\mu}, \sigma}_{A}$ and similar for the arm $B$.
In order to calculate the density matrix of the polarization entangled state $\rho$ we trace the density matrix $\tilde{\rho} = \ket{\phi}\bra{\phi}$ of the state in \eqref{eq_PP}  over the signal and idler angular frequencies and obtain
\begin{equation}
\label{eq_den}
\rho = 1/\mathcal{N} \int \hspace{-1.5ex} \int d\omega^{\prime} d \omega^{\prime\prime} 
\vphantom{}_{A}\negmedspace \bra{\omega^{\prime}}\vphantom{}_{B}\negmedspace\bra{\omega^{\prime\prime}} \tilde{\rho}\ket{\omega^{\prime\prime}}_{B}  \ket{\omega^{\prime}}_{A},
\end{equation}
in which $\mathcal{N}$ is a normalization constant. Using the relation $\vphantom{}_{A}\negmedspace \braket{\omega^{\prime} | \omega_{\mu}, \sigma}_{A}= \delta(\omega^{\prime}-\omega_{\mu})\ket{\sigma}_{A}$ and similar for the arm $B$, we can rewrite the density matrix  of the polarization entangled state in \eqref{eq_den} as
\begin{align}
\label{eq_rho_final}
\rho = &\alpha \ket{HV}_{AB\ AB}\hspace{-0.4ex}\bra{HV}
+\mathcal{D}(\tau) \ket{VH}_{AB\ AB}\hspace{-0.4ex} \bra{HV} \nonumber \\
+&\mathcal{D}^{*}(\tau) \ket{HV}_{AB\ AB}\hspace{-0.4ex} \bra{VH}
+\beta\ket{VH}_{AB\ AB}\hspace{-0.4ex} \bra{VH}
\end{align}
in which $\alpha$ and $\beta$ are the diagonal elements of the density matrix, the sum of which is normalized to unity. The complex-valued off-diagonal elements, $\mathcal{D}(\tau)$ and its complex conjugate $\mathcal{D}^{*}(\tau)$, quantify the amount of entanglement in the created state, which we call the  \emph{degree of polarization entanglement}, shortly $\mathcal{D}$-parameter   \cite{Zhukovsky2012}. This parameter is highly dependent on $\tau$ and the state characteristics can be manipulated by changing the relative delay between signal and idler. 

The elements of the density matrix in \eqref{eq_rho_final} are determined by the spectral properties of the PDC emission and can be evaluated via $\alpha =1/\mathcal{N} \int \hspace{-1ex} \int d\omega^{\prime} d \omega^{\prime\prime} |g(\omega^{\prime}, \omega^{\prime\prime})|^{2}$,  $\beta = 1/\mathcal{N}\int \hspace{-1ex} \int d\omega^{\prime} d \omega^{\prime\prime} |h(\omega^{\prime\prime}, \omega^{\prime})|^{2}$ as well as
\begin{align}
\label{eq_D}
\mathcal{D}(\tau) = 1/\mathcal{N}\int \hspace{-1ex} \int d\omega^{\prime} d \omega^{\prime\prime} \textrm{e}^{i(\omega^{\prime}-\omega^{\prime\prime}) \tau}h(\omega^{\prime\prime}, \omega^{\prime})g^{*}(\omega^{\prime}\omega^{\prime\prime}).
\end{align}
We emphasize that the trace in \eqref{eq_den} generally causes some mixedness and therefore, the density matrix cannot be written in terms of a pure polarization entangled state. Moreover, the purity of the state in \eqref{eq_rho_final} is greatly dependent on the temporal characteristics of the $\mathcal{D}$-parameter that are governed by the properties of the underlying PDC process \cite{Zhukovsky2012, Fang2014,Lim2016}.

%----------------------------------------------------------------------------------
%                       Sample
%----------------------------------------------------------------------------------
\begin{figure}[b]
\centering
\includegraphics[width = 0.5\textwidth]{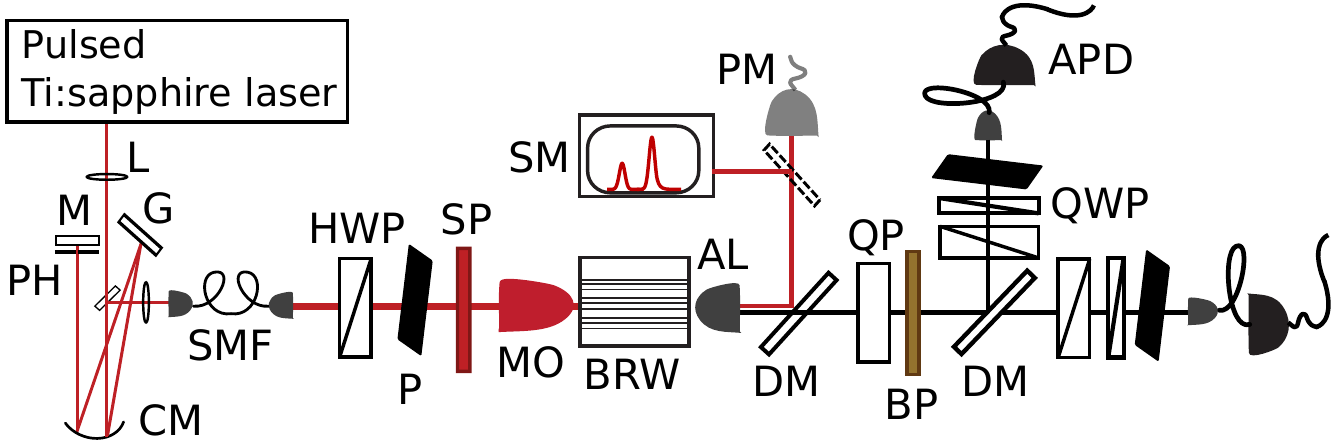}
\caption{\label{fig:exp} Setup for creating  and characterizing the polarization entanglement from a BRW. Abbreviations: AL = aspheric lens, APD = avalanche photo-diode,  BP = band pass filter,  CM = curved mirror,  DM = dichroic mirror, G = grating, HWP = half-wave plate, L = lens, M = mirror, MO = microscope objective,  P = polarizer, PH = pinhole, PM = power meter, QP = quartz plate, QWP = quarter-wave plate,   SM = spectrometer,  SMF = single mode fiber, SP = short pass filter. }
\end{figure}

In our experiment shown in Fig.~\ref{fig:exp} we use a pulsed laser as the pump for the PDC process, whose spectral properties (\unit{767.6}{\nano\meter} central wavelength, \unit{0.8}{\nano\meter} bandwidth) are controlled with a grating setup. After spatial mode control with a single mode fiber, the proper power and  polarization of the pump beam are chosen. The pump pulses are coupled into our BRW with a 100x microscope objective and collimated afterwards with a high numerical aperture aspheric lens having a \unit{3}{\milli\meter} focal length.  The investigated BRW with the same layer structure as in Refs \cite{Guenthner2015, B.Pressl2015, Laiho2016} has \unit{4}{\micro\meter} ridge width, is  \unit{1.87}{\milli\meter} long and its front facet is anti-reflection coated for the pump wavelengths. Signal and idler are emitted into the fundamental total internal reflection modes, whereas the pump is a higher-order spatial mode called Bragg mode.
Behind the BRW, the pump beam with an average power of about \unit{250}{\micro\watt} is separated from the emitted signal and idler beams with a dichroic mirror.  Thereafter, a spectral filter with \unit{40}{\nano\meter} bandwidth and a central wavelength of \unit{1535.2}{\nano\meter}, corresponding to the PDC degeneracy wavelength, is employed to limit the spectral  extent of the emitted photon pairs in order to suppress background light. The PDC emission is then sent to a dichroic mirror that is centered at the degeneracy and has nominally a \unit{7}{\nano\meter} wide step between transmission and reflection bands  for both polarizations.  In the two output arms we use a combination of half- and quarter-wave plates as well as sheet polarizers for setting the desired polarizations for the tomographic reconstruction  of the density matrix.  Finally, we couple the transmitted light into single mode fibers and use commercial, gated avalanche photo-detectors having a quantum  efficiency of about $0.2$ for counting the coincidences and singles at a measurement rate of \unit{1.9}{\mega\hertz} and acquisition time of \unit{2}{\minute}.

The pump power  is selected such that the coincidences-to-accidentals ratio (CAR) is approximately $9.5(5)$. We extract the CAR from the measurement of the coincidence and single count rates that are about \unit{4\hspace{-0.5ex}}{/\second} and \unit{870\hspace{-0.5ex}}{/\second}, respectively, when one of the the arms $A$ and $B$ passes light in $H$-polarization and the other one in $V$-polarization \cite{Guenthner2015}. A value as high as possible for the CAR is desired in order to suppress the higher photon-number contributions  in the PDC emission and spurious counts that cause an undesired background on the measured coincidences. It is also strong evidence of the concurrent production of photons in signal and idler beams, i.e. the pair production.

We implement a relative temporal delay between signal and idler with  a \unit{0.82}{\milli\meter} thick quartz plate \cite{Grice1997}, which according to our earlier experiments \cite{Laiho2016} is necessary for compensating the group delay between signal and idler.  We verify the path length difference along the retarder's fast and slow axis with the help of a spectrophotometer having a spectral range of \unit{0.5-2}{\micro \meter} (Perkin-Elmer, Lambda 19). We place the retarder between two crossed polarizers such that the input polarization is rotated in the retarder due to its birefringence. The strong wavelength dependence of the retardance causes fringes in the system's transmission. This approach yields a relative temporal difference of $\Delta\tau =$ \unit{25.9(4)}{\femto \second} between its fast and slow axes near \unit{1550}{\nano\meter}.

%----------------------------------------------------------------------------------
%               Results
%----------------------------------------------------------------------------------

We reconstruct the density matrices of the generated states from 36 correlation measurements between the two output paths of the dichroic mirror. The created state is projected onto all polarization combinations of  $H/V$, $D^{+}/D^{-} = (H\pm V)/\sqrt{2}$ and $R/L=(H\pm \imath V)/\sqrt{2}$.  For the state tomography we employ standard optimization algorithms with the subtraction of accidental counts \cite{James2001}. We further note that although achromatic  components are used, a polarimeter study reveals small deviations from the desired polarization within the broad spectral band of signal and idler.

\begin{figure}[t]
\centering
\includegraphics[width = 0.5\textwidth]{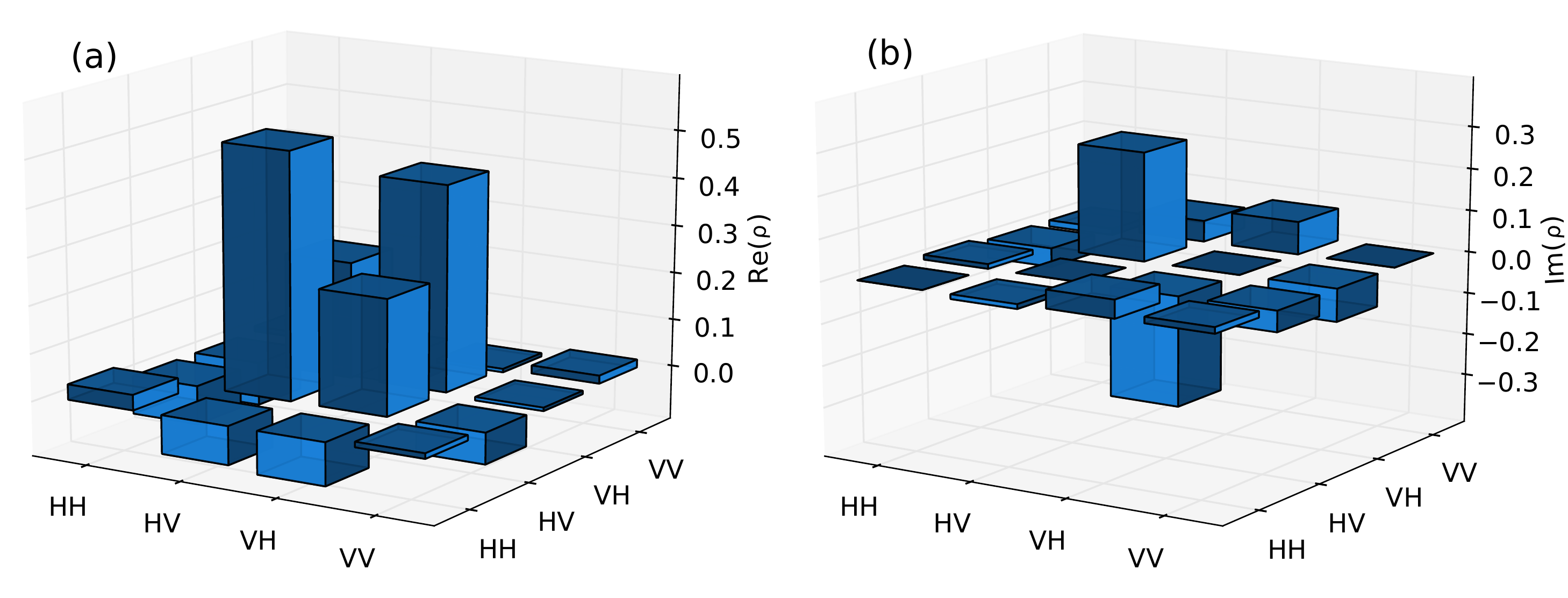}
\caption{\label{fig:result1} (a) Real and (b) imaginary part of the reconstructed density matrix for the case (i). See Data File 1 for underlying values. The two main diagonal elements of the density matrix  take the values of $0.52(1)$ and $0.43(1)$. The undesired background in the remaining diagonal elements is about $0.05$.  The $\mathcal{D}$-parameter  takes the value of $\ 0.243(13)\pm\imath \cdot0.259(12)$ corresponding to the weights of the main off-diagonal elements.}
\end{figure}

The relative time delay between signal and idler is controlled with the orientation of the fast axis of the  quartz plate  or by removing it from the setup. In total we can measure the density matrix at three temporal delays of (i) 0, (ii) +$\Delta\tau $ and (iii) $-\Delta\tau$.
In Fig. \ref{fig:result1} we show the reconstructed density matrix for the case (i). We clearly notice that the main diagonal elements of the density matrices have slightly different weights that can be explained by a slight mismatch in the central wavelengths of the broadband filter and the dichroic mirror. 
We further notice that there is a rather large noise background in many  elements of the reconstructed density matrix. It is most probably caused by background light such as fluorescence from the BRW and imperfect optics distorting the polarization of light. Also the rather low CAR is another sign of the existence of these spurious counts.
 However, most importantly, we see that the $\mathcal{D}$-parameter exhibits both real and imaginary parts, which means that the intrinsic group delay of signal and idler is non-negligible.

\begin{figure}[t]
\centering
\includegraphics[width = 0.4\textwidth]{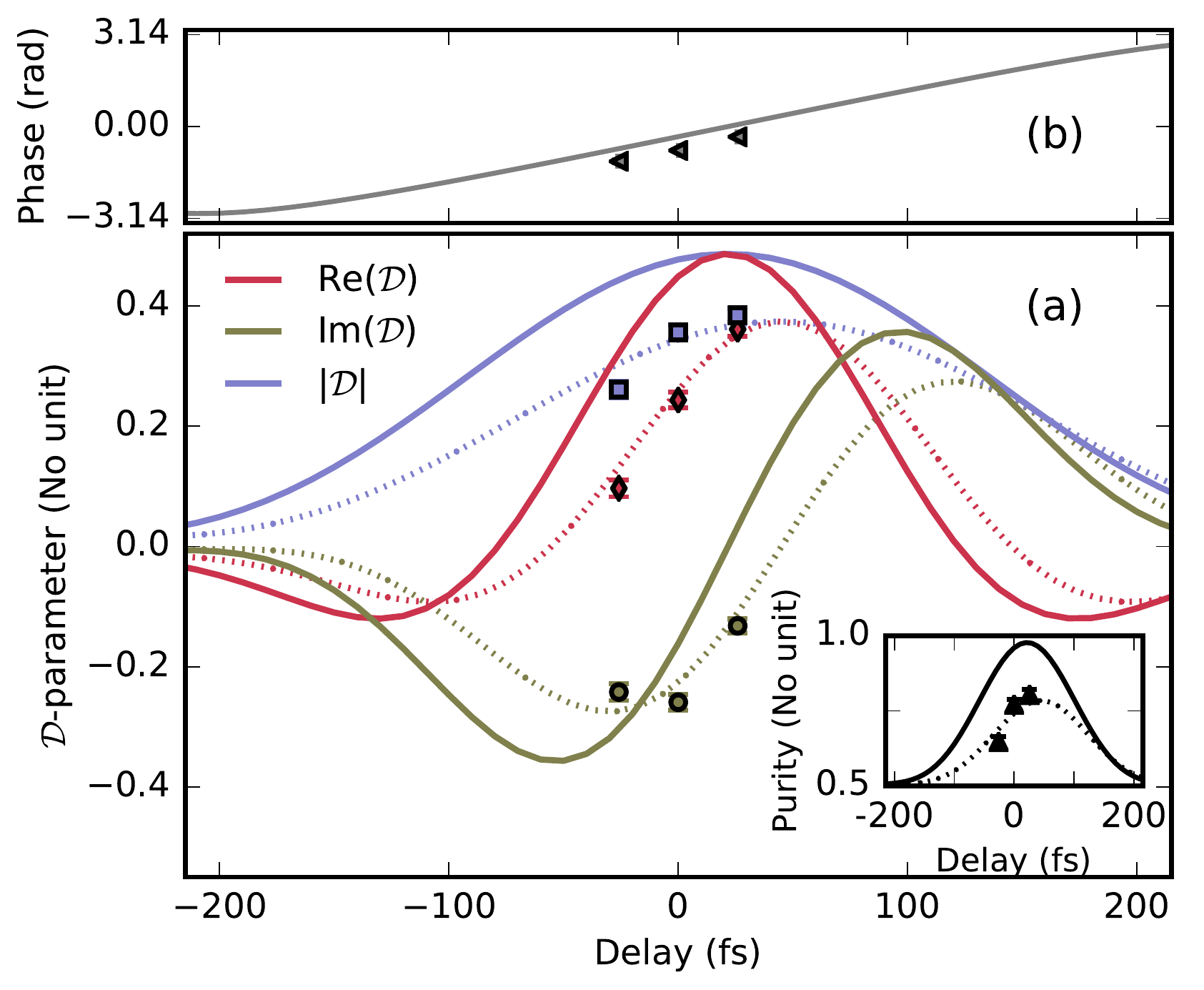}
\caption{\label{fig:result2} (a) Predicted  values for the real and imaginary parts as well as for the absolute value of the $\mathcal{D}$-parameter vs.~the temporal delay. The  measured values are marked with diamonds, circles and squares, respectively. The dotted lines illustrate the corresponding degradation of the theoretical curves due to the experimental imperfections. The inset shows the expected (solid line), due to the imperfections degraded (dotted line) and measured (symbols) purity extracted via $\textrm{Tr}\big( \rho^2 \big)$. (b) The predicted phase (solid line) extracted via $\arg{\big(\mathcal{D}\big)}$ vs. the temporal delay together with the measured values (symbols).}
\end{figure}

Next, we use the quartz plate to manipulate the $\mathcal{D}$-parameter, whereas the diagonal elements of the density matrix vary only slightly. In case (ii) we compensate  the relative temporal  delay between signal and idler to suppress the imaginary part of the off-diagonal terms in the density matrix. From the raw coincidence counts measured for the state tomography we extract  the maximal visibilities of 0.78(3), 0.68(3) and 0.68(3) in $H$/$V$, $D^{+}$/$D^{-}$ and R/L basis, respectively. We achieve a  $\mathcal{D} $-parameter with the value  of $0.361(12) \pm \imath \cdot 0.132(13) $. This parameter can also  become fully imaginary close to the delay used in case (iii), in which case we extract the value of $0.097(15)\pm \imath \cdot 0.242(14)$. 
In order to examine this time-dependence in more detail, we illustrate in Fig. \ref{fig:result2}(a) the obtained $\mathcal{D}$-parameters with respect to the temporal delay between signal and idler together with \eqref{eq_D}, which is  calculated with the JSA in Fig.~\ref{fig:polent}(b).  
The inset shows that the purity of the created state drops strongly from its maximum value for temporal delays of a few hundred femtoseconds. In Fig. \ref{fig:result2}(b) we extract the phase of the $\mathcal{D}$-parameter, which exhibits  a linear behavior.

The highest values for the $\mathcal{D}$-parameter and  purity are reached as expected when the relative temporal delay between signal and idler is compensated, and the degree of polarization entanglement should be real-valued.  Otherwise the state is mixed, finally exhibiting no coherence, in other words, the off-diagonal elements of the density matrix vanish. Nevertheless, within this time window we can tune the phase of the $\mathcal{D}$-parameter by changing the delay between signal and idler. Our theoretical predictions can qualitatively verify the expected trend, however, the measured values are somewhat decreased.  
We take these imperfections into account empirically by shifting the theoretical prediction for $\mathcal{D}(\tau)$ in time with the offset clearly visible in Fig.~\ref{fig:result2}(b) between the experiment and theory
and by scaling down its amplitude, which is diminished from the prediction with the
background-free theory as already indicated by the measured raw visibilities in $D^{+}/D^{-}$ and R/L bases. 
Finally, we find a good agreement with the theory and experiment also shown in Fig. \ref{fig:result2}(a).  We summarize the achieved values for the purity, fidelity  and concurrence  \cite{Hill1997}  in Table \ref{tab:purity}.

\begin{table}[!tb]
 \centering 
\vspace{2ex}
 \begin{tabular}{cccc}
 \hline
Case&Purity & Fidelity & Concurrence \\
\hline
(i)&0.77(2)&93.0(6)&70(3)\\
(ii)&0.80(2)&93.2(6)&75(3)\\
(iii)&0.65(2)&92.2(7)&50(3)
 \end{tabular}
\caption{\label{tab:purity} Extracted values for the purity, fidelity with the simulated states from \eqref{eq_rho_final} evaluated via $\textrm{Tr}\big({\sqrt{\rho^{1/2} \rho_{\textrm{Exp}} \rho^{1/2}}}\big)$ with $\rho_{\textrm{Exp}}$ being the experimentally reconstructed density matrix, and concurrence for the cases (i)-(iii).}
 \end{table}

%----------------------------------------------------------------------------------
%                Conclusions
%----------------------------------------------------------------------------------

To conclude, we studied the temporal characteristics of the polarization entangled states in terms of the $\mathcal{D}$-parameter. These states were created by splitting the broadband, pulsed PDC emission from our BRW with a dichroic mirror and by manipulating the relative delay between signal and idler prior to that. Our results illustrate the necessity of an accurate temporal control when creating these states via PDC  emitted from BRWs. We have shown that the group delay difference between signal and idler is non-negligible, affecting the $\mathcal{D}$-parameter in an uncontrolled manner. However, its phase can easily be tuned if the relative delay between signal and idler is changed, excluding the need for modifying the state afterwards in the individual paths of the two separate parties. Our results deliver insight into the internal phase of the PDC emission and emphasizes the possibilities for the state manipulation directly at the source.

%----------------------------------------------------------------------------------
%                Acknowledgements
%----------------------------------------------------------------------------------

This work was supported by the FWF through project no. I-2065-N27, the DFG Project no. SCHN1376/2-1, the ERC project \textit{EnSeNa} (257531) and the State of Bavaria. We thank  A. Wolf and S. Kuhn for assistance during sample growth and fabrication.

\end{document}